\documentclass[conference]{IEEEtran}
\usepackage{cite}
\usepackage{amsmath,amssymb,amsfonts}
\usepackage{graphicx}
\usepackage{eso-pic}
\usepackage[draft]{hyperref}
\begin{document}

\AddToShipoutPictureBG*{\AtPageLowerLeft{\put(\LenToUnit{\dimexpr(\paperwidth-\textwidth)/2\relax},\LenToUnit{10mm}){\parbox[b]{\textwidth}{\footnotesize This work has been submitted to the IEEE for possible publication. Copyright may be transferred without notice, after which this version may no longer be accessible.}}}}

\title{Optical Motion Sensing for Piezoelectric-Resonator-Based Power Converters}

\author{\IEEEauthorblockN{Liyang Jin\IEEEauthorrefmark{1},
Zhiyang Cao\IEEEauthorrefmark{2},
Jun Ji\IEEEauthorrefmark{1},
Liyan Zhu\IEEEauthorrefmark{2}, and
Linbo Shao\IEEEauthorrefmark{1}\IEEEauthorrefmark{3}}
\IEEEauthorblockA{\IEEEauthorrefmark{1}Department of Electrical and Computer Engineering, Virginia Tech, Blacksburg, USA}
\IEEEauthorblockA{\IEEEauthorrefmark{2}Center for Power Electronics Systems, Virginia Tech, Blacksburg, USA}
\IEEEauthorblockA{\IEEEauthorrefmark{3}Center for Quantum Information Science and Engineering (VTQ), Virginia Tech, Blacksburg, USA}
\IEEEauthorblockA{\{liyangjin, tsaozy, junji, liyanz, shaolb\}@vt.edu}
}

\maketitle

\begin{abstract}
Piezoelectric components present a promising alternative to conventional magnetics for miniaturized, high-frequency power converters. 
While high-Q piezoelectric resonators have demonstrated high efficiencies and large power densities in prototype high-frequency dc-dc converters, 
their optimal operating conditions are sensitive to temperature fluctuations and load variations. 
Therefore, robust closed-loop control is essential for practical deployment.
Electrical approaches for sensing motional current may compromise the electrical isolation and electromagnetic interference (EMI) immunity inherently provided by the mechanical nature of these resonators. 
Here, we present an optical motion sensing approach for piezoelectric power converters. 
The optical sensing system comprises a laser, fiber optics, and a photodetector. 
The proposed approach uses optical interference to monitor the mechanical displacement in real time. 
The resulting optical signal carries both the amplitude and phase of the mechanical motion, which are necessary for closed-loop control. 
We experimentally characterize and validate this optical motion sensing approach using a 45-kHz piezoelectric resonator driven by a GaN half bridge with input voltages up to 45~V.
We further experimentally demonstrate closed-loop control using the proposed optical motion sensing for a 99-kHz isolated piezoelectric converter. 
These initial results establish a viable pathway toward optically closed-loop control of piezoelectric power converters.
\end{abstract}

\begin{IEEEkeywords}
Optical sensing, piezoelectric converter, closed-loop control
\end{IEEEkeywords}

\section{Introduction\label{sec:intro}}
Piezoelectric components~\cite{Yao2025transducers, Yao2026tuson, Galanko2019led, Qian2026natcomm}, offering thin profiles, high power densities, and high Q factors, are compelling alternatives to conventional magnetic components in power converters~\cite{Wang2022tpel, Daniel2024apec, Naval2026tpel, Sanz2003apec, Boles2022tpel, Boles2023tpel, Stolt2024tpel, Boles2023ieej}.
Unlike conventional inductors and transformers, which scale poorly to small volumes and high frequencies, 
piezoelectric resonators (PRs) store energy in mechanical vibrations to achieve high power densities and low losses at MHz switching frequencies. 
Despite these physical advantages, implementing closed-loop control for high-frequency PR-based dc-dc converters remains challenging.
Efficient power conversion requires operation in the inductive regions of PRs, achieving zero-voltage and zero-current switching (ZVS, ZCS), and reducing looping current. 
To achieve these, the controller must have the phase information of the current in the motional branch of the PR, namely the motional current $I_m$.
Researchers have proposed various approaches to detect the zero-crossing points of $I_m$, by using shunt resistors or diodes~\cite{Yang2020tpel}, adding additional electrodes to PRs~\cite{Cao2026tpel}, implementing a real-time algorithm~\cite{Bigot2024iet}, or deriving them from the voltage waveform across PRs sampled by high-speed analog-to-digital converters (ADCs)~\cite{Piel2021compel, Touhami2026ojpel}.
These approaches pioneered the closed-loop control of PR-based power converters. Nevertheless, a more direct measurement of $I_m$ is of interest. Fiber-optic interferometers have been used to characterize the vibration of piezoelectric elements~\cite{DaSilva2011spie}. More broadly, optical sensing offers electrical isolation and electromagnetic interference (EMI) immunity in power electronics~\cite{Yang2026apex}.

This work presents an optical motion sensing approach for piezoelectric power converters. 
Using optical interference, our approach directly measures the physical displacement of the PR with both phase and amplitude information.
The optical fiber provides electrical isolation, and the resonator needs no additional electrodes or contacts.
We experimentally validate the optical motion sensing mechanism using a 45-kHz PR driven by a GaN half bridge.
The detected optical signal keeps a fixed phase relative to the current measured by a current probe, and its modulation depth is proportional to the current amplitude.
We further demonstrate closed-loop control of optically sensed piezoelectric power converters.
The optical signal feeds the same type of simple event-driven controller as in~\cite{Cao2026tpel}.

\begin{figure}[h]
    \centering
    \includegraphics[]{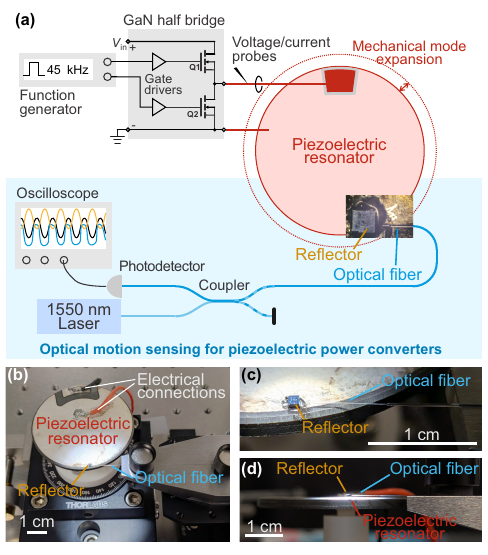}
    \caption{Optical motion sensing for piezoelectric converters. (a) Schematic of the setup with a GaN half-bridge circuit, a piezoelectric resonator, and an optical motion sensing system. (b)--(d) Photos of the experimental setup.}
    \label{fig:optical-sensing-system}
\end{figure}

\section{Architecture of Optically Sensed Piezoelectric Converter}
Fig.~\ref{fig:optical-sensing-system}(a) illustrates the schematic of a PR driven by a power circuit and sensed by the proposed optical system.
The PR is driven by a GaN-based half bridge with a dc input voltage $V_{\mathrm{in}}$, and the gate signal is controlled by a function generator.
The half bridge drives the PR near its resonant frequency.
The optical sensing system comprises a continuous-wave laser (Santec, TSL770), which emits narrow-linewidth laser light in the telecommunication wavelength band of 1550~nm,
an InGaAs photoreceiver (New Focus, 1811-FC, 125-MHz bandwidth),
a 90:10 fiber directional coupler that splits light into the excitation path and detection path, 
and a cleaved fiber positioned facing an optical reflector mounted on the PR.
Although the PRs used in this work vibrate at 45 and 99~kHz, the 125-MHz bandwidth of the photoreceiver leaves room for MHz resonators.
All optical fibers are Corning SMF-28, a standard telecom fiber.
For sensor characterization, an oscilloscope is used to capture the output signal (proportional to the light intensity) of the photoreceiver. 
We use the smooth sidewall of a chip die as an optical reflector. 
The photos of the PR and the optical-fiber-based sensor are shown in Figs.~\ref{fig:optical-sensing-system}(b)--\ref{fig:optical-sensing-system}(d).

In the demonstration of closed-loop control in Section~\ref{sec:closed-loop}, the photoreceiver output passes through a band-pass filter (BPF) and a comparator to a control board based on a field-programmable gate array (FPGA). Also, the light directly hits the sidewall of the PR, i.e., no reflector is mounted on the PR.

\section{Working Principle of Optical Motion Sensing}
\begin{figure*}[ht]
    \centering
    \includegraphics[]{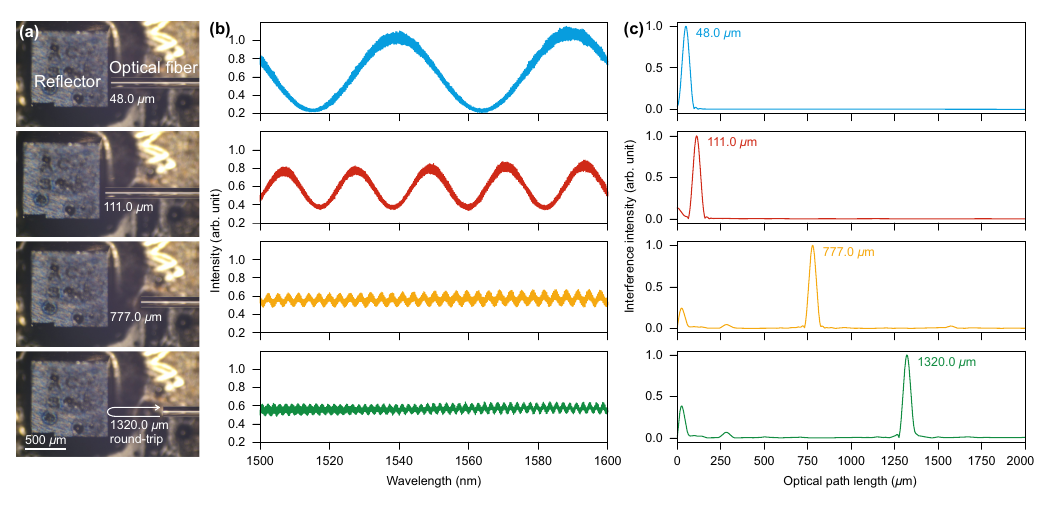}
    \caption{Principle of optical motion sensing. (a) Microscope photos showing the sensing optical fiber at different distances from the reflector, labeled with the round-trip optical path length. (b) Measured optical spectra at corresponding distances. (c) Optical path length spectra extracted from the measured optical spectra.}
    \label{fig:optical-sensing-principle}
\end{figure*}

The proposed optical motion sensing approach relies on optical interference
between the cleaved fiber end and the optical reflector (Fig.~\ref{fig:optical-sensing-principle}(a)).
The cleaved fiber end and the reflector form a low-finesse Fabry--Perot interferometer.
Depending on the optical wavelength, the reflected beams interfere constructively or destructively, producing a wavelength-dependent variation in the received optical intensity.
To characterize this interference across different fiber-to-reflector distances, we measure the optical spectra by capturing the received optical
intensity as the laser wavelength is swept from 1500 to 1600~nm
(Fig.~\ref{fig:optical-sensing-principle}(b)). 
A shorter distance between the
fiber end and the reflector yields a larger spectral modulation period, known
as the free spectral range (FSR). 
Applying a Fourier transform to
the measured optical spectra \cite{Thomas2025optica} yields the optical path
length spectra (Fig.~\ref{fig:optical-sensing-principle}(c)), where each peak
identifies the corresponding optical path length present in the
interferometer.

In general, the received optical intensity spectrum is given by
\begin{equation}
    I = I_{0} \left( 1 + \gamma \cos\left(\frac{2\pi}{\lambda} L + \phi_o\right) \right),
    \label{eq1}
\end{equation}
where $I_0$ is the average optical power received at the photodetector,
$\gamma \in [0,1]$ is the interference visibility, $\lambda$ is the optical
wavelength, $L$ is the round-trip optical path length between the fiber end and the reflector, and $\phi_o$ is the interference
phase offset.
The interference visibility $\gamma$, which determines the
signal-to-noise ratio, is set by the optical intensities reflected
from the two interfaces: the cleaved fiber end and the reflector.
Specifically, $\gamma$ reaches $1$ when the two reflected optical intensities are
equal. 
For the measurements shown in
Fig.~\ref{fig:optical-sensing-principle}(b), the optical intensity coupled
back from the reflector is lower than that from the fiber end. A larger fiber-to-reflector distance reduces the reflected light
recaptured by the fiber. 
Consequently, the visibility $\gamma$ decreases as the distance increases.

When the PR is driven by a half bridge, the optical reflector moves with the mechanical mode, resulting in distance variations between the reflector and the fiber end. 
For real-time optical sensing, we park the laser at a fixed wavelength so that the varying fiber-to-reflector distance directly modulates the received optical intensity. 
A sinusoidal mechanical motion yields an optical path length of $L = L_0 + 2d\cos(2\pi f_m t + \phi_d)$,
 where $L_0$ is the static round-trip path, $d$ is the displacement amplitude at the sensed position, the factor of 2 accounts for the round trip, $f_m$ is the vibration frequency, and $\phi_d$ is the mechanical phase. 
Substituting $L$ into Eq.~\eqref{eq1} gives the detected optical intensity waveform
\begin{equation}
    I(t) = I_{0} \left( 1 + \gamma \cos\left(\beta \cos(2\pi f_m t + \phi_d) + \phi_o\right) \right),
    \label{eq2}
\end{equation}
where $\beta = 4\pi d/\lambda$ is the phase modulation index (optical modulation depth).
We tune the laser wavelength to place the operating point near the middle of an interference fringe, $\phi_o \approx \pm\pi/2$, where the response to displacement is largest, and the sign of the fringe slope sets the polarity of the signal.
Equation~\eqref{eq2} matches the waveforms measured in Section~\ref{sec:demo}.

In the closed-loop control demonstration (Section~\ref{sec:closed-loop}), we apply the BPF at the fundamental vibration frequency $f_m$ to obtain a clean sinusoidal feedback signal. At the quadrature operating point ($\phi_o = \pi/2$), expanding Eq.~\eqref{eq2} via the Jacobi--Anger identity gives
\begin{equation}
\begin{aligned}
    I(t) = I_0 \Big[ 1 &- 2\gamma J_1(\beta) \cos(2\pi f_m t + \phi_d) \\
    &+ 2\gamma J_3(\beta) \cos(6\pi f_m t + 3\phi_d) - \cdots \Big].
\end{aligned}
\label{eq3}
\end{equation}
Here, $J_n(\beta)$ is the $n$th-order Bessel function of the first kind. 
Therefore, applying the BPF at the fundamental vibration frequency $f_m$ yields a sinusoidal feedback signal with an amplitude of $2I_0 \gamma J_1(\beta)$. This amplitude is linear in $\beta$ only for $\beta \lesssim 1$.
For larger motion, the ratio of the third harmonic to the fundamental, $H_3/H_1 = J_3(\beta)/J_1(\beta)$, which depends only on $\beta$ and not on $I_0$, $\gamma$, or $\phi_o$, could be used to determine the displacement amplitude of the PR.

\begin{figure}[h]
    \centering
    \includegraphics[width=0.9\linewidth]{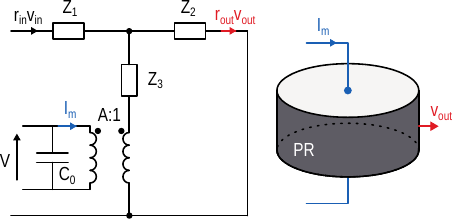}
    \caption{Electromechanical model of disk-shaped piezoelectric resonators. }
    \label{fig:em-model}
\end{figure}

As described in Section~\ref{sec:intro}, $I_m$ is the physical quantity that must be measured to implement closed-loop control.
It is thus of interest to recover the $I_m$ information from the detected optical signals.
Fig.~\ref{fig:em-model} illustrates an electromechanical model of a disk-shaped PR~\cite{Forrester2023jmems}, where $Z_x$ ($x$ = 1 to 3) denotes the mechanical impedances, $r_{in} = 0$ and $v_{in} = 0$, $r_{out}$ is the radius of the component, $v_{out}$ is the vibration velocity at the outer edge, $A$ represents the conversion ratio between electrical and mechanical energy, $V$ is the excitation voltage, and $C_0$ is the static capacitance.
Ideally, the phase of $I_m$ exactly matches the phase of the vibration velocity on the side of the PR, $v_{out}$.
Since the displacement is $x = \int {{v_{out}}{\rm d}t}$, the displacement lags $I_m$ by $90^\circ$.
The optical signal therefore leads or lags $I_m$ by $90^\circ$, depending on the slope of the interference fringe at the operating point.
This fixed offset is compensated by an analog phase-shifting circuit (including the BPF) or a digital delay unit, whereas the measured amplitude of the terminal current follows from the optical modulation depth $\beta$ (Fig.~\ref{fig:linearity}).

\section{Demonstration of Optical Motion Sensing\label{sec:demo}}
We experimentally validate the optical motion sensing using the setup shown in Fig.~\ref{fig:optical-sensing-system}(a). 
In this experimental demonstration, we use a commercial ultrasonic transducer as the PR.
The PR features a radial mechanical mode at 45~kHz, and thus we drive it near this resonant frequency to efficiently excite the mechanical motion. 

\begin{figure*}[!t]
    \centering
    \includegraphics[width=\linewidth]{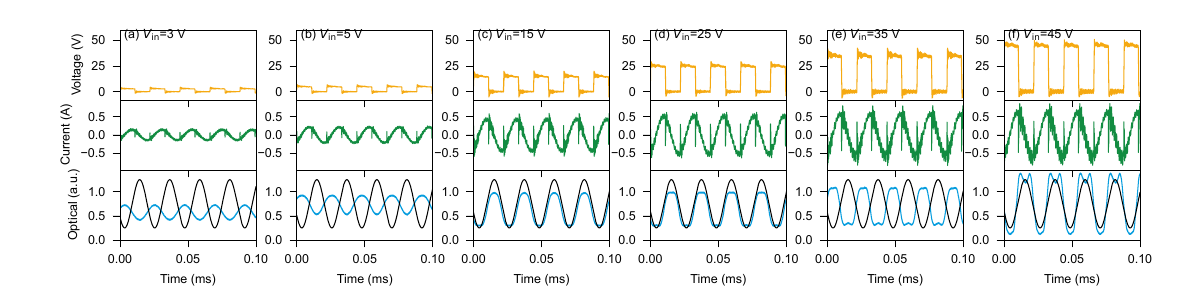}
    \caption{Optically sensed signal of the PR at different input voltages from (a) 3~V to (f) 45~V. Yellow curves (first row) are the voltage at the input of the PR, green curves (second row) are the terminal current flowing into the PR, blue curves (third row) are the raw detected optical signals, and black curves (third row) are phase reference waveforms recovered by fitting the raw signals to Eq.~\eqref{eq2}.}
    \label{fig:optical-sensing-at-diff-voltage}
\end{figure*}

Fig.~\ref{fig:optical-sensing-at-diff-voltage} shows the measured optical signal waveforms, the applied switch voltage, and the measured driving current waveforms at different input voltages. 
The current probe measures the terminal current, which includes spikes at each switching edge (green curves in the second row).
The optical signal (blue curves in the third row) shows no such spikes, which is critical for robust closed-loop control.
For a given fringe slope, the optical signal keeps a fixed phase relative to the measured current at all input voltages, which provides the timing information for the switching sequence~\cite{Boles2021tpel, Cao2026tpel}.
The fundamental component of the optical signal, selected by the BPF, can thus serve as the feedback signal in closed-loop control.

\begin{figure}[th]
    \centering
    \includegraphics[]{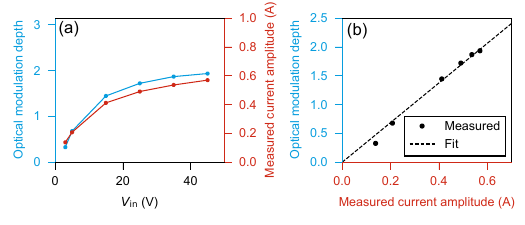}
    \caption{Quantitative comparison between optical motion sensing and electrical current measurement. (a) Extracted optical modulation depth ($\beta$, blue curve, left axis) and measured current amplitude (red curve, right axis) as functions of the input voltage $V_{\text{in}}$ from 3 to 45~V. (b) Optical modulation depth versus measured current amplitude (black circles) with a proportional fit (dashed line), showing the proportionality between the optical and electrical motion sensing.}
    \label{fig:linearity}
\end{figure}

The raw optical amplitude grows with the input voltage (Figs.~\ref{fig:optical-sensing-at-diff-voltage}(a)--(c)) until higher harmonics, mainly the third-order term $2 I_0 \gamma J_3(\beta)$, distort the waveforms (Figs.~\ref{fig:optical-sensing-at-diff-voltage}(d)--(f)).
To quantitatively evaluate the sensing linearity across the full operating range, we extract the optical modulation depth $\beta$ by fitting the measured optical waveforms in Fig.~\ref{fig:optical-sensing-at-diff-voltage} to Eq.~\eqref{eq2} and compare it with the measured driving current amplitude (Fig.~\ref{fig:linearity}). 
As shown in Fig.~\ref{fig:linearity}(a), as $V_{\mathrm{in}}$ increases from 3 to 45~V, the extracted $\beta$ increases from 0.33 to 1.93,
corresponding to a mechanical displacement amplitude $d = \beta\lambda/(4\pi)$ from 41 to 238~nm, 
closely tracking the measured current amplitude from 0.14 to 0.57~A. 
Both curves become sublinear above about 5~V, likely due to thermal drift of the resonant frequency at large driving voltage.
Although the third-order harmonic term $2I_0 \gamma J_3(\beta)$ in Eq.~\eqref{eq3} distorts the raw optical waveforms at higher drive levels (Figs.~\ref{fig:optical-sensing-at-diff-voltage}(d)--(f)), the extracted $\beta$ remains proportional to the measured current amplitude, as shown in Fig.~\ref{fig:linearity}(b).
Because the motional current is expected to be proportional to the vibration velocity, and thus to the displacement amplitude at a fixed frequency, this proportionality indicates that the optical sensor accurately captures the mechanical motion of the piezoelectric resonator.
Because $\beta$ is proportional to the measured current amplitude, the harmonic ratio $H_3/H_1$ of the optical signal could give the current amplitude directly, without waveform fitting.

\section{Closed-Loop Control with Optical Feedback \label{sec:closed-loop}}

\begin{figure}
    \centering
    \includegraphics[width=0.9\linewidth]{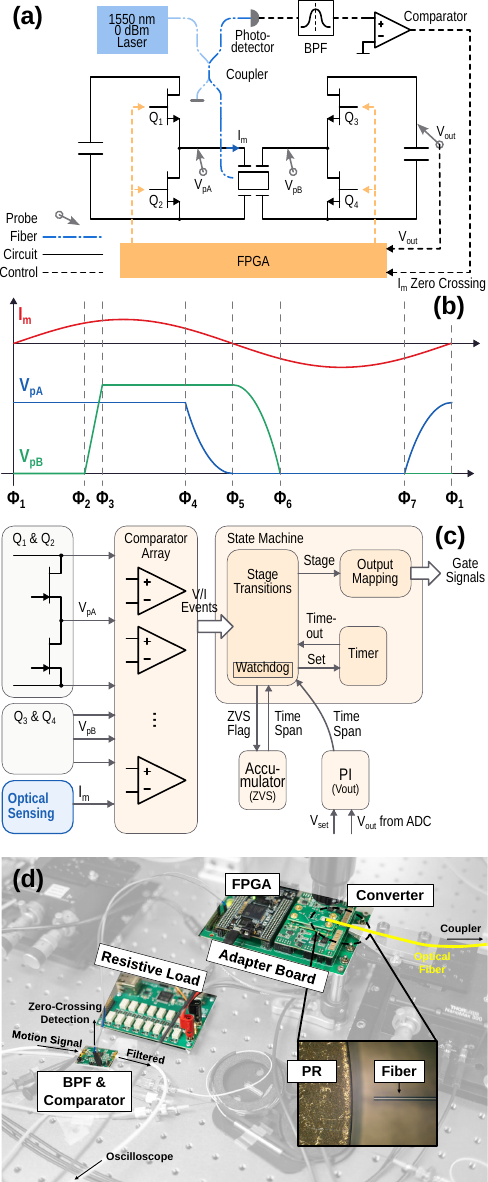}
    \caption{Demonstration of a closed-loop-controlled isolated piezoelectric converter with optical motion feedback. (a) System-level block diagram of the converter. (b) Idealized steady-state operation waveform. (c) Control system implemented on an FPGA. (d) Experimental setup.}
    \label{fig:converter-exp}
\end{figure}

We further propose a closed-loop control system for isolated piezoelectric converters with optical feedback, as shown in Fig.~\ref{fig:converter-exp}. 
The structure of the converter is illustrated in Fig.~\ref{fig:converter-exp}(a), where both the primary and secondary sides of the PR are connected to the middle points of GaN half bridges and driven in the sequence shown in Fig.~\ref{fig:converter-exp}(b)~\cite{Boles2022jestpe}.
 Vibration of the PR is picked up by the fiber, translated into an electrical signal, and then sent to the controller.
 A BPF is inserted to filter out higher-order harmonics of the optical signal and low-frequency drift of the mechanical system.
In our experimental setup (Fig.~\ref{fig:converter-exp}(d)), the PR is fabricated based on APC item 2470 (diameter $D=23.0~\rm{mm}$, thickness $H=2.7~\rm{mm}$, resonant frequency $f_r \approx 99~ \rm{kHz}$).
The converter operates at $V_{in} = \mathrm{20}$~V, $V_{out} = \mathrm{26}$~V, and $P_{out} = \mathrm{0.34}$~W. In this demonstration, the optical fiber points directly at the edge of the PR without an additional reflector.

$\Phi_1$ to $\Phi_7$ in Fig.~\ref{fig:converter-exp}(b) denote the seven transition points of the switching sequence. To minimize current circulation and realize ZCS, $\Phi_1$ and $\Phi_5$ should be triggered at the zero-crossing points of $I_m$, which are detected by the zero-crossing points of the optical signal in our proposed system. $\Phi_3$ and $\Phi_6$ are triggered when another group of comparators detects that $V_{pB}$ passes over $V_{out}$ or under zero, respectively. $\Phi_4$ and $\Phi_7$ are controlled by a binary proportional-integral (PI) controller (i.e., a digital accumulator) to guarantee that $V_{pA}$ reaches zero exactly at the negative zero-crossing point of $I_m$, and reaches $V_{in}$ at the positive zero-crossing point. Lastly, $\Phi_2$ is determined by another PI loop to regulate the voltage conversion ratio. The system is implemented on an FPGA as described in Fig.~\ref{fig:converter-exp}(c). Detailed information regarding the binary PI loop and the state machine can be found in~\cite{Cao2026tpel}.

\begin{figure}[h]
    \centering
    \includegraphics[width=0.9\linewidth]{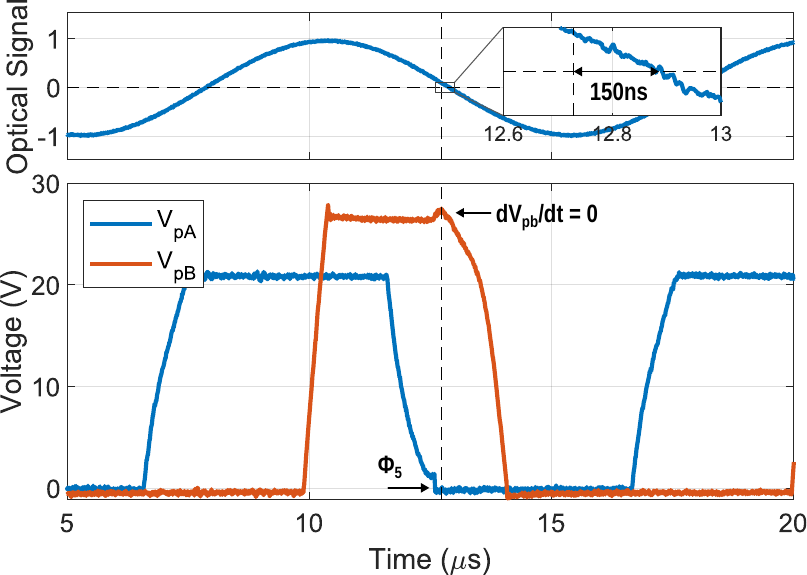}
    \caption{Recorded steady-state operation waveform. The optical signal obtained from the photodetector is filtered by the BPF and normalized for clarity.}
    \label{fig:converter-steaty-state-waveform}
\end{figure}

The steady-state operation waveforms of $V_{pA}$, $V_{pB}$, and the filtered optical signal are recorded and shown in Fig.~\ref{fig:converter-steaty-state-waveform}.
Although the motional current cannot be directly measured, its zero crossings can be derived from the node voltages.
While the $V_{pB}$ node floats, $I_m$ alone charges the node capacitance, so $I_m$ crosses zero where $dV_{pB}/dt=0$~\cite{Cao2026tpel}.
The dashed line in Fig.~\ref{fig:converter-steaty-state-waveform} marks one such point.
The filtered optical signal crosses zero about 150~ns later, which we attribute to the delays of the BPF and the photoreceiver.
To better demonstrate this delay, the FPGA triggers $\Phi_1$ and $\Phi_5$ 300~ns before the optical zero crossing in experiment, which is realized by delaying the preceding opposite crossing by $T/2 - 300~\rm{ns}$, where $T$ is the switching period.

\section{Conclusion and Future Work}
This work demonstrates optical motion sensing for piezoelectric power converters and its use as the feedback signal for closed-loop control.
The approach provides a direct, electrically isolated measurement of the mechanical vibration, without additional electrodes or contacts on the resonator.
Future work includes locking the optical operating point and extending the approach to MHz resonators.

\section*{Acknowledgment}
The authors used AI models (Claude Fable~5.2, GPT~6.1 Astra, and their sub-agents) to revise the entire manuscript.
Code generated by multiple AI models (GPT~6.1 Astra, Claude Opus 4 and 5) was used for automatic equipment control and data processing, and the authors tested and confirmed its results.
The original draft and all claims are from the authors.
The authors take full responsibility for the accuracy and integrity of the manuscript.

\bibliographystyle{IEEEtran}
\bibliography{linbo_ref}

\end{document}